\documentstyle[12pt]{article}

\begin{document}
\setcounter{page}{1}

\pagestyle{plain} \vspace{1cm}
\begin{center}
\Large{\bf Can Quantum Gravitational Effects Manifest themselves at Large Distances?}\\
\small
\vspace{1cm} {\bf Kourosh Nozari} \quad and \quad {\bf Behnaz Fazlpour}\\
\vspace{0.5cm} {\it Department of Physics,
Faculty of Basic Sciences,\\
University of Mazandaran,\\
P. O. Box 47416-1467, Babolsar, IRAN} \\
e-mail: knozari@umz.ac.ir

\end{center}
\vspace{1.5cm}

\begin{abstract}
Consider a proposed model of the universe with $\hbar$ much greater
than its well-known value of $10^{-34} Js$. In this model universe,
very large objects can show quantum behaviors. In a scenario with
large extra dimensions, $\hbar$ can attains very large values
depending on the dimensionality of spacetime. In this letter, we
show that although conventional thinking indicates that quantum
gravitational effects should manifest themselves only at very small
scales, in actuality quantum gravitational effects can manifest
themselves at large scales too. We use the generalized uncertainty
principle with a non-zero minimal uncertainty in momentum as our
primary input to construct a mathematical framework for our proposal.

\end{abstract}

\section{Motivation}
The realm of quantum theory is very small scale, which usually
consists atomic and subatomic levels. Gravity is a long range
interaction and acts on ordinary scales as well as very large scales
such as stellar and galactic ones. Usually we expect that quantum
effects manifest themselves at very small scales. It is hard to
accept that gravity as an interaction governing the large scale
structures, can show quantum behavior at large distances. Here we
are going to show that actually, gravity can show quantum behavior
in large distances. This idea is in contradiction with classical
belief which accepts that quantum effects manifest themselves at
very small scales. Recently, {\it El Naschie} has proposed a
framework for investigation of such a fundamental problem[1]. He has
indicated that " There remains the major problem of relating a
theory developed for cosmological scales to problems on the scale of
high energy elementary particles with wave particle dualism,
tunneling effects and quantum entanglement." He then provides an
elegant discussion of the problem using $T$-duality in $E$-infinity
space. In our opinion, the formalism presented here will provide a
suitable framework to deal with such a fundamental problem. To
formulate our idea, we use extra dimensional scenarios and also the
generalized uncertainty principle which consists of an absolute
non-zero minimum uncertainty in momentum. We provide some clarifying
examples to show in what extent the objects can manifest quantum
effects under the given conditions.
\section{An Effective $\hbar$}
We begin our argument with the following question: can $\hbar$ be
varying? To answer this question we consider the following
generalized uncertainty relation which has been motivated from
string theory and other approaches to quantum gravity[2-4]
\begin{equation}
\Delta x \Delta p\geq \hbar\bigg(1+\alpha^{2} l_{p}^{2}\frac{(\Delta
p)^2}{\hbar^2}\bigg),
\end{equation}
where we have considered the absolute minimum of uncertainties to
avoid the appearance of expectation values in the right hand side.
The second term on the right hand side has its origin on the quantum
effects of gravitation. If one tries to set a high precision
position measurement of an electron, he should use very energetic
photons. But in this situation one has to consider the gravitational
interaction of photon and electron. In fact, a high energy photon
causes spacetime fluctuation. Taking this point into account, one
finds the second term of the right hand side as an extra uncertainty
due to gravitational interaction of photon and electron. As a
result, this generalized uncertainty relation consists of a natural
cut off on the order of Planck length. Note that $\alpha$ is a
string theory parameter of the order unity. Comparison with standard
uncertainty relation $\Delta x \Delta p\geq \hbar$ shows that we can
define
\begin{equation}
\hbar_{eff}={\hbar}\bigg(1+\alpha^{2} l_{p}^{2}\frac{(\Delta
p)^2}{\hbar^2}\bigg).
\end{equation}
The extra term on the right hand side of (1) is important at high
momentum regime, where one can write approximately $\Delta p\sim p$.
Therefore, one can obtain the following generalization
\begin{equation}
\hbar_{eff}={\hbar}\bigg(1+\alpha^{2} l_{p}^{2}\frac{
p^2}{\hbar^2}\bigg).
\end{equation}
In ordinary quantum mechanics, de Broglie principle is given by
$\bar{\lambda}=\frac{\hbar}{p}$. Now equation (3) can lead to the
following generalization of de Broglie principle
\begin{equation}
\bar{\lambda}_{eff}=\frac{\hbar_{eff}}{p}=\frac{\hbar}{p}\bigg(1+\alpha^{2}
l_{p}^{2}\frac{ p^2}{\hbar^2}\bigg).
\end{equation}
This generalization will affect fundamental arguments of quantum
theory. For example, the entire argument of wave mechanics such as
wave propagation, wave broadening  and other domains such as
coherent states of quantum mechanical systems should be re-examined
within this framework[5-7]. Note that equation (4) can be
interpreted in another fashion. We can write it in the following
form
\begin{equation}
\bar{\lambda}_{eff}=\frac{\hbar}{p_{eff}}=\frac{\hbar}{p}(1+\alpha^{2}
l_{p}^{2}\frac{ p^2}{\hbar^2}\bigg).
\end{equation}
This point of view leads to "modified dispersion relation" which can
be written as follows
\begin{equation}
\frac{1}{p_{eff}}=\frac{1}{p}\bigg(1+\alpha^{2} l_{p}^{2}\frac{
p^2}{\hbar^2}\bigg).
\end{equation}
Modified dispersion relations have some signature in ultra-high
cosmic ray showers which are under serious investigation[8].\\
Up to this point, we have been satisfied that $\hbar$ can be varying
with momentum. This idea has its very basic notion in the string
theoretical considerations.

\section{ $\hbar$ in Models with Extra Dimensions}
In this section, we argue that in scenarios with extra
dimensions[9,10],  $\hbar$ can attains very large values relative to
its $4$-dimensional counterpart. In $4$-dimensions, one can write
$$m_{p}l_{p}=\hbar,$$
while in $4+d$ dimensions(we use Arkani-Hamed, Dimopoulos and Dvali
model of extra dimensions[9]) this statement generalizes to
$$M_{f}L_{f}=\hbar,$$
where $M_{f}$ and $L_{f}$ are Planck mass and Planck length in model
universe with extra dimensions respectively[11]. The relation
between $4$-dimensional Planck length and the extension of extra
dimensions $R$(we suppose all extra dimensions have the same
extension) is given by
\begin{equation}
m_{p}^{2}=M_{f}^{2+d}R^{d}.
\end{equation}
Therefore, we can write
\begin{equation}
M_{f}=\frac{\hbar}{L_{f}}=\frac{\hbar_{eff}}{l_{p}}=\frac{\hbar_{eff}}{\hbar}m_{p}
\quad \Longrightarrow \quad
M_{f}^{2}=(\frac{\hbar_{eff}}{\hbar})^{2}M_{f}^{2+d}R^{d}.
\end{equation}
This relation can led us to the following result
\begin{equation}
\hbar_{eff}=\frac{\hbar}{(M_{f}R)^{\frac{d}{2}}}.
\end{equation}
The extension(radius) $R$ of these extra dimensions, for $M_{f}\sim
TeV$, typically lies in the range from $1 mm$ to $10^{3} fm$ for $d$
from $2$ to $7$. Suppose the case with $d=3$, that is, with only
three extra dimensions. In this case $R=10 nm =10^{-8}m $. With
$M_{f}=1TeV=1.6\times10^{-7}J$, we find
$$\hbar_{eff}\sim 10^{-11} Js.$$
In a model universe with this value of Planck constant, de Broglie
relation $\lambda_{eff}=\frac{\hbar_{eff}}{mv}$ shows that a
particle with detectable wave properties and with usual speed can
have very large mass. For Example, suppose that $\lambda=10^{-14}m$
which is in the range of detectors resolution and $v=100m/s$. We
find $m=10 kg$. Such a large mass has quantum effects in a universe
with $\hbar_{eff}\sim 10^{-11}Js$.\\
Thus far, we have been satisfied that quantum effects are not
restricted to very small masses depending on the value of $\hbar$.
In the next section we show that gravity can manifest quantum
behavior in large distances.

\section{Quantum Gravity Effects in Large Distances}
Spacetime at short distances has a noncommutative structure. This
noncommutativity can be addressed in the generalized uncertainty
principle. Recently we have shown that spacetime noncommutativity
and the generalized uncertainty principle when are applied to the
issue of black hole thermodynamics, give the same results[12]. Since
uncertainty principle is the foundation of quantum theory and
gravity induces uncertainty, one has to incorporate gravitational
uncertainty from the very beginning of the quantum theory
formulation. There are several possibilities to write these
generalized uncertainty principles. The most general form of
generalized uncertainty principle can be expressed as[13]
\begin{equation}
\Delta x\Delta p\geq\hbar\bigg(1+\beta^{2}\frac{(\Delta
x)^2}{l_{p}^{2}}+\alpha^{2}l_{p}^{2}\frac{(\Delta
p)^2}{\hbar^{2}}+\gamma\bigg).
\end{equation}
Since we are dealing with absolute minimum of uncertainties, we
suppose that $\gamma=0$. The resulting generalized uncertainty
principle has minimal non-zero uncertainty both in position and
momentum. One possible choice is the case where $\alpha=0$. Then we
find
\begin{equation}
\Delta x \Delta p\geq\hbar\bigg(1+\beta^{2}\frac{(\Delta
x)^{2}}{l_{p}^{2}}\bigg).
\end{equation}
In this case we have non-zero minimal uncertainty only in momentum.
Quantum gravitational effects are important only in the limit where
the second term in the right hand side of (11) becomes comparable
with the first term, that is where
\begin{equation}
\beta\frac{\Delta x}{l_{p}}\approx 1\Longrightarrow \Delta x\approx
\frac{l_{p}}{\beta}.
\end{equation}
Since usually $\beta\ll 1$, this statement shows that quantum
gravitational effects can manifest themselves in very large
distances. Physically this is the case since generalized uncertainty
principle in the form of (11) contains quantum gravitational effects
in the limit of high $\Delta x$. Therefore, we conclude that quantum
effects of gravitation can be revealed in large distances as well as
very small distances. Note that in $E$-infinity theory of {\it El
Naschie} such a result can be obtain by admitting and tuning of
dimensions in a concrete way. A similar argumentation but without
relating its to extra dimensional scenarios has been pointed in
[14]. It seems that there is a correspondence between large/small
distance behavior of quantum gravitational effects. This idea can be
examined in the next generation of accelerators, specially the Large
Hadronic Colider(LHC)[15].

\section{Summary}
In this letter we have argued that quantum gravitational effects can
manifest themselves at large distances as well as small distances.
We have found a generalization of  $\hbar$  and also de Broglie
principle to show the possibility of having quantum effects for
large masses. We have used string theory generalized uncertainty
principle and ADD scenario of extra dimensions for formulation of
our idea. Our proposal has the ability to be examined in the next
generation of Hadronic Coliders at CERN. Actually, possible
verification of extra dimensions proposed by ADD model and also
possible detection of $TeV$ black hole remnants in LHC and ILC will
provide indirect experimental test of our conjecture[16,17].\\

\end{document}